\documentclass[12pt]{iopart}
\usepackage{epsfig}
\usepackage{amssymb}

\def\half{\frac{1}{2}}

\def\e{\epsilon}
\def\eqref#1{(\ref{#1})}

\begin{document}
\title{Condensation transition in zero-range processes with diffusion}

\author{E. Levine, D. Mukamel and G. Ziv}

\address{ Department of Physics of Complex Systems, Weizmann
Institute of Science, Rehovot, Israel 76100.}

\begin{abstract}
Recent studies have indicated that the coarse grained dynamics of
a large class of traffic models and driven-diffusive systems may
be described by urn models. We consider a class of one-dimensional
urn models whereby particles hop from an urn to its nearest
neighbor by a rate which decays with the occupation number $k$ of
the departure site as $\left(1+b/k\right)$. In addition a
diffusion process takes place, whereby all particles in an urn may
hop to an adjacent one with some rate $\alpha$. Condensation
transition which may take place in this model is studied and the
($b,\alpha$) phase diagram is calculated within the mean field
approximation and by numerical simulations. A driven-diffusive
model whose coarse grained dynamics corresponds to this urn model
is considered.
\end{abstract}

\section{Introduction}
\label{sec:intro}

Ordering and condensation transitions in one-dimensional systems
far from thermal equilibrium have been studied extensively in
recent years \cite{Mukamel00,Evans00,GunterRev}. It has been
repeatedly demonstrated that unlike systems in thermal
equilibrium, driven one-dimensional systems whose dynamics does
not obey detailed-balance can be ordered even when the dynamics is
local and noisy.

By making a correspondence between phase separation in
one-dimensional systems and condensation in urn models, two
mechanisms which allow for phase separation in driven systems have
been suggested. Urn models are simple lattice models defined on a
ring geometry, where each site can either be vacant or occupied by
one or more particles. The first mechanism is described in terms
of the Zero Range Process (ZRP) \cite{Evans00,ZRP}. In this model
particles hop between nearest neighbor lattice sites with rates
$\omega_k$ which depend only on the number of particles $k$ at the
departure site. If the rates decay to zero in the large $k$ limit,
or if the rates decay slowly enough to a finite value, a
condensation transition takes place, whereby a single lattice site
becomes macroscopically occupied as the density is increased
beyond a critical value. It has been suggested that the
coarse-grained dynamics in a broad class of one-dimensional driven
models can be described by a ZRP with rates which at large $k$
decay as $\omega_k = \omega_\infty\left(1+b/k\right)$
\cite{Kafri02A,Kafri03}. In this case phase separation takes place
only if $b>2$.

A second mechanism is described in terms of the of the Chipping
Model (CM) \cite{Majumdar98,Rajesh02}. The dynamics of this model
involves two processes: {\em chipping}, where a single particle
hops to a nearest neighbor site at a constant rate $\omega$ ; and
{\em diffusion}, where all particles in a site hop together to an
adjacent site with rate $\alpha$. This model can be viewed as a
ZRP with constant hopping rate ($b=0$), extended to include
diffusion processes as well. Mean-field analysis of this model
indicates that this model exhibits a condensation transition at a
critical density. This result remains valid beyond the mean-field
approximation as long as the chipping process is symmetric
\cite{Majumdar98}. It has also been shown that if the chipping
process is biased either to the left or to the right, no
condensation transition takes place \cite{Rajesh02}.

Recently it has been suggested that the coarse-grained dynamics of
certain cellular-automata traffic models can be modeled by the
asymmetric CM \cite{Levine03}. Within this picture, the
coarse-grained evolution of traffic models is described in terms
of domain dynamics, which essentially involves asymmetric chipping
and diffusion processes. It has thus been concluded that no phase
separation transition should be expected in this class of models,
and that jamming phenomena take place as a broad crossover process
rather than via a sharp phase transition. On the other hand, there
exist models which may be related to traffic, in which the rate of
chipping a particle from a domain is not constant, but rather
depends on the domain size. Such an example is the Bus Route model
\cite{BRM}, where an approximate description assigns hopping rates
to buses that decay as a function of the distance to the next bus
ahead. It is thus of interest to combine the features of the
chipping model and the ZRP processes and consider urn modes which
exhibit chipping and diffusion processes with occupation dependent
chipping rates.

In this paper we consider in details a class of urn models which
incorporate both diffusion and chipping processes with occupation
($k$) dependent chipping rates of the form $\omega_k = 1+b/k$. In
Section 2 we introduce the generalized ZRP and analyze its
behavior both within mean-field approximation and by numerical
simulations. In Section 3 we consider a simple driven diffusive
model, and show that its coarse grained dynamics is described by
this generalized ZRP. Conclusions and summary are presented in
Section 4.

\section{Zero-Range Processes with diffusion}
\label{sec:zrpcm}

The generalized zero-range process is defined on a lattice of $M$
sites, with periodic boundary conditions, occupied by $N = \phi M$
particles. The dynamics is defined through the rates by which two
nearest neighbor sites containing $k$ and $m$ particles,
respectively, exchange particles:
\begin{eqnarray*}
\fl \mbox{diffusion:} \qquad(k,m)
\mathop{\longrightarrow}^{\alpha}(k+m,0) \\
\fl \mbox{chipping:}\qquad
(k,m)\mathop{\longrightarrow}^{q\omega_m}(k+1,m-1) \qquad,\qquad
(k,m)\mathop{\longrightarrow}^{(1-q)\omega_k}(k-1,m+1)\,.
\label{eq:gcm_rates}\end{eqnarray*}
 where the rate $\omega_k$ depends on the
number of particles in the departure site,
\begin{equation}
\omega_k = 1 + \frac{b}{k}\;.
\end{equation}
Thus the model is specified by three dynamical parameters,
$\alpha, b$ and $q$, and by the average occupancy $\phi$. The
limit $\alpha=0$ corresponds to a ZRP with no diffusion; the limit
$b=0$ recovers the chipping model. The parameter $q$ controls the
spatial bias, with $q=1/2$ being the symmetric case.

\subsection{Mean-Field analysis}
We first analyze the model within mean-field approximation, where
correlations between sites are neglected. Let $p_k$ be the
probability of a given site to be occupied by $k$ particles. The
mean-field evolution equations for $p_k$ are given by
\begin{eqnarray}
\label{eq:zrpcm} \fl \frac{\partial p_0}{\partial t} &=&
\alpha(1-p_0)^2+\omega_1 p_1(1-p_0)-(\lambda-\omega_1 p_1)\,p_0
\\ \fl \nonumber
 \frac{\partial p_n}{\partial t} &=&
-2\,\alpha\,p_n(1-p_0) + \alpha\sum_{k=1}^{n-1}p_kp_{n-k}
+\left[\lambda
(p_{n-1}-p_n)+\omega_{n+1}p_{n+1}-\omega_{n}p_n\right] \;,
\end{eqnarray}
with $n \geq 1$ and $\lambda=\sum_{k=1}^\infty \omega_k p_k$.

To solve these equations in the steady state, where all time
derivatives vanish, let us define a generating function $g(s) =
\sum_{n=1}^\infty p_n s^n/ n$. Multiplying the $n$th equation
in~\eqref{eq:zrpcm} by $s^n$ and summing over $n \geq 1$ one
obtains
\begin{eqnarray}
\label{eq:zrpcm_gs}  \alpha s^2g'(s)^2 -2\,\alpha s(1-p_0)g'(s)
+(1-s)(1-\lambda\,s)g'(s) \\ \nonumber  \qquad\qquad\qquad + \;\;
b\frac{1-s}{s}g(s) -\lambda p_0 (1-s) + \alpha (1-p_0)^2 = 0\;.
\end{eqnarray}
It is assumed that at criticality $p_k$ has the asymptotic large
$k$ behavior $p_k \sim 1/k^{\tau}$. One is interested in
evaluating $\tau$ as a function of $\alpha$ and $b$. For the model
to exhibit a condensation transition one needs $\tau>2$. Within
the mean-field approximation the model exhibits condensation
transition already at $b=0$. We therefore expect such transition
to take place for any $b \geq 0$ (an assertion which has been
checked numerically), and hence assume in the following that
$\tau>2$. This form of $p_k$ implies that $g(s)$ is well-defined
for $|s|<1$, and is singular at $s=1$. Taking $s \equiv
e^{-\epsilon}$, with $\epsilon>0$, we find that for non-integer
$\tau$ the singular part of $g(s)$ has the following behavior for
$\epsilon \ll 1$,
\begin{equation}
g(s) \sim  \int_1^\infty \frac{e^{-\epsilon n}}{n^{\tau+1}}dn \sim
\epsilon^{\tau}\,,
\end{equation}
where for the purpose of extracting the leading singularity we
replace the sum by an integral. For integer $\tau$ the singularity
of $g(s)$ is of the form $\epsilon^\tau\log\epsilon$. In the
following analysis we assume for convenience that $\tau$ is
non-integer. The results derived in this analysis apply to the
integer $\tau$ case as well. We thus make the anzats
\begin{equation}\label{eq:zrpcm_anz}
g(s)=\left(a_0+a_1\epsilon+a_2\epsilon^2+\ldots\right)+\epsilon^\tau\,
\left(b_0+b_1\epsilon+b_2\epsilon^2+\ldots\right)+\ldots\,,
\end{equation}
where the terms in the first brackets represent the regular part
of $g(s)$, while the terms in the second brackets correspond to
the singular part, with $\tau$ being its leading power. Clearly
non-leading singularities with powers $\tau^\prime>\tau$ can exist
in this expansion as well. In the following we proceed by
inserting this anzats into \eqref{eq:zrpcm_gs}, and systematically
solving this equation order by order in $\epsilon$. From the
equations for terms of order $0, 1, 2$ and $\tau$ one readily
derives expressions for both the critical occupancy $\phi_c$ and
$p_0$ in terms of $\lambda$ :
\begin{equation}
\label{eq:zrpcm_phic}
 \phi_c = \frac{1-\lambda}{2\alpha}\qquad ,
\qquad
p_0=1-\frac{\lambda}{b}+\frac{(1-\lambda)^2}{4\,\alpha\,b}\,.
\end{equation}
Note that non-leading singular terms cannot contribute to terms of
these orders, and they thus need not be considered.

\begin{figure}[t]
\centerline{\epsfig{file=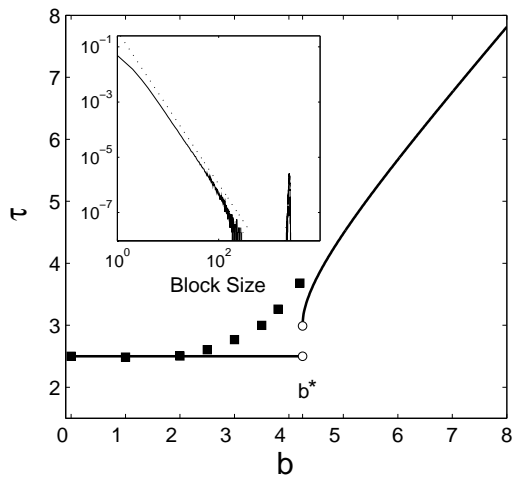,width=7truecm}} \caption{The
exponent $\tau$ as a function of $b$ for $\alpha=0.2$. Solid line
is given by the solution to the mean-field equations. Boxes are
results of numerical simulations of the mean-field dynamics with
$M=1000, \phi=3$. An estimation for $\tau$ is obtained by fitting
$p_k$ to $k^{-\tau}$ over the range $10<k<100$. An example for
such a fit in the case $b=3$ is given in the inset.
\label{fig:zrpcm_mf_tau}}
\end{figure}

In order to determine $\lambda$ and $\tau$ one needs to
distinguish between two cases :
\begin{description}
\item[({\it i} )] $ \tau > 3$ ---  One proceeds by considering the
orders $3, 4$ and $\tau+1$ terms in \eqref{eq:zrpcm_gs}. Since
$\tau>3$, non-leading singularities do not contribute to these
terms as well. A straightforward calculation yields
%
%
\begin{equation}
\label{eq:zrpcm_lambda_gt3} \hspace{-1cm} \lambda=\frac{2
\alpha+b+1-\sqrt{\gamma}}{3}\,,
\end{equation}
\begin{equation}
\label{eq:zrpcm_tau_gt3}  \hspace{-1cm} \tau = 9b
\left[3b-\sqrt{3}\left(4\gamma + 3b^2 + 3 -6b
-4(2\alpha+b+1)\sqrt{\gamma}\right)^{1/2}\right]^{-1}\;,
\end{equation}
where $\gamma = 4 \alpha\,(\alpha + b + 1)+(b-2)^2$. Inserting
\eqref{eq:zrpcm_lambda_gt3} into \eqref{eq:zrpcm_phic} one gets
the critical occupancy $\phi_c$ in terms of the model parameters.
This solution holds for values of $b$ larger than $b^*$, defined
by $\tau(b^*) = 3$. Note that in the limit $b \to \infty$ one has
$\tau \to b$ . This is a result of the fact that at large values
of $b$ diffusion becomes rare and the ZRP result is approached.
\item[({\it ii} )] $2 < \tau < 3$ --- In this interval we find
that for $b<b^*$ (for which $\tau<3$) $\tau$ becomes independent
of $b$, with $\tau=5/2$. To analyze this case we consider the
${2\tau-2}$ term in \eqref{eq:zrpcm_gs}. This term is given by
$\alpha \tau^2 b_0^2 \e^{2\tau-2}$, which does not vanish since by
definition $b_0 \neq 0$. In this $\tau$ interval non-leading
singular terms in $g(s)$ cannot contribute to this term. To see
this, let $\tau^\prime > \tau$ correspond to a such sub-leading
term. Its leading contribution to~\eqref{eq:zrpcm_gs} is of the
order $\e^{\tau^\prime+1}$, which is of higher order than
$\e^{2\tau-2}$. This analysis implies that $2\tau-2$ has to be an
integer. Within this $\tau$ interval, the only possibility is
$\tau = 5/2$.

The parameter $\lambda$, and thus the critical occupancy, cannot
be determined perturbatively in this $\tau$ interval. It is
determined by the two boundary conditions imposed on
\eqref{eq:zrpcm_gs}, namely $g(0)=0$ and $g(1)=(\lambda-1+p_0)/b$,
with $p_0$ given by \eqref{eq:zrpcm_phic}. Integrating
\ref{eq:zrpcm_gs} numerically, one identifies the value of
$\lambda$ which satisfies the boundary conditions.
\end{description}

\begin{figure}[t]
\centerline{\epsfig{file=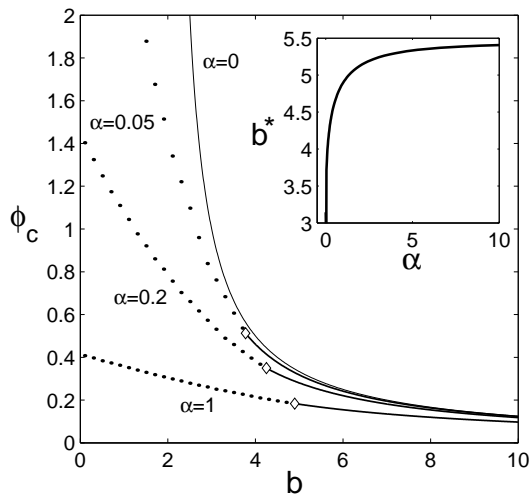,width=7truecm}}
\caption{Mean-field critical occupancy $\phi_c$ as a function of
$b$ for several values of $\alpha$. Solid lines are given by
\eqref{eq:zrpcm_phic} and \eqref{eq:zrpcm_lambda_gt3} for
$\tau>3$. Dots are obtained numerically from \eqref{eq:zrpcm_gs}
for $\tau<3$, as explained in the text. The $\alpha=0$ line is
given by the ZRP solution, $\phi_c=1/(b-2)$. The $\diamondsuit$
symbols indicate the value of $b^*$ on each curve. At $b=0$ the
critical density is given by the CM solution
$\phi_c=\sqrt{1+1/\alpha}-1$. The value of $b^*$ as a function of
$\alpha$, as obtained in mean-field, is given in the inset.
\label{fig:zrpcm_mf_phic}}
\end{figure}

In summary, we find that within the mean-field approximation
$\tau=5/2$ for $b<b^*(\alpha)$, while $\tau$ is a continuous
function of $b$ for $b>b^*(\alpha)$. The exponent $\tau$ exhibits
a discontinuity at $b^*(\alpha)$ (see
figure~\ref{fig:zrpcm_mf_tau}). The critical occupancy $\phi_c$ is
given as a function of $b$ in figure~\ref{fig:zrpcm_mf_phic}. For
$b>b^*(\alpha)$ the curve is deduced from
\eqref{eq:zrpcm_lambda_gt3} and \eqref{eq:zrpcm_phic}, while for
$b<b^*(\alpha)$ it is obtained numerically by finding the value of
$\lambda$ for which the solution of \eqref{eq:zrpcm_gs} satisfies
the appropriate boundary conditions. One observes that although
the exponent $\tau$ exhibits a discontinuity at $b^*$, the
critical occupancy is a continuous monotonically decreasing
function of $b$. Note that increasing either $\alpha$ or $b$
results in decreasing the critical occupancy. This is due to the
fact that condensation is favored by both the diffusion process
and by the size-dependency of the chipping. The value of $b^*$ as
a function of $\alpha$ is depicted in the inset of this figure.

\begin{figure}[t]
\centerline{\epsfig{file=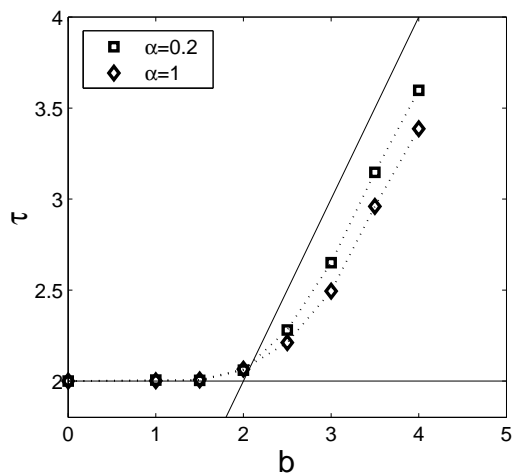,width=7truecm}} \caption{The
exponent $\tau$ as a function of $b$, obtained from numerical
simulations. Solid lines are $\tau=2$ corresponding to asymmetric
CM ($b=0$), and $\tau=b$ corresponding to ZRP ($\alpha=0$).
Simulations were performed on systems of size $M=1000$ and
occupancy $\phi=3$. \label{fig:zrpcm_oned_tau}}
\end{figure}

The $\tau(b)$ curve found above corresponds to the mean-field
approximation in the thermodynamic limit. It is of interest to
examine how this behavior is manifested in finite systems. This
could give a useful insight for the model with nearest-neighbor
exchange, where no analytic results are available, and where one
has to resort only to numerical simulations of finite systems. To
this end we carried out numerical simulations of the model, with a
modified dynamics for which the mean-field equations yield the
correct steady state. This modified dynamics is defined by the
same rates as those of the original model, except that the two
sites exchanging particles in each dynamical move are chosen at
random, and are not necessarily nearest neighbors. The results
obtained for $\tau$ from a simulation of a system of size $M=1000$
are presented in figure~\ref{fig:zrpcm_mf_tau}. It is observed
that the qualitative behavior derived in the previous section is
recovered. Namely, $\tau$ assumes the value $5/2$ for small $b$,
while it continuously varies with $b$ at large $b$. In the
vicinity of the critical parameter $b^*$, large crossover effects
dominate the dynamics, and the numerical curve seems to deviate
from the analytical one. This is a finite size effect, and one
needs far larger systems in order to recover the true asymptotic
exponent $\tau$ in this region, as given by the exact solution
found above. We carried out similar calculations for larger
systems and found that indeed the trend is to move towards the
analytical curve.

\subsection{Numerical simulations}

In the case of nearest-neighbor exchange no analytical solution
for the steady-state distribution function is available, and one
has to resort to numerical simulations.  The exponent $\tau(b)$
for the totally asymmetric case ($q=0$), as extracted from studies
of systems of size $M=1000$, is given in
figure~\ref{fig:zrpcm_oned_tau}. It is readily seen that the
numerical results are consistent with $\tau = 2$ at small $b$ and
with a continuously varying $\tau$ at large $b$.
This indicates that as in the CM ($b=0$) the generalized model
does not exhibit a condensation transition for small $b>0$.
Condensation is obtained only above a certain threshold of the
parameter $b$. The precise nature of the curve in hard to
determine numerically, due to the finite size effects which
dominate the dynamics in the vicinity of $b^*$, as for the
mean-field dynamics. Thus, for example, accurate determination of
$b^*$ is not possible, although it seems close, or equal, to $2$.
It is also not clear from the simulations whether or not $\tau(b)$
is discontinues at $b^*$.

We also carried out numerical studies of the symmetric case,
$q=\half$. For $b=0$ it is known that $\tau=5/2$, as in mean-field
\cite{Majumdar98}. We find that $\tau$ remains locked at $5/2$ for
sufficiently small $b$, while $\tau$  varies continuously above
this value for large $b$.

\section{Corresponding one-dimensional driven diffusive model} \label{sec:dds}

We now demonstrate that the generalized zero-range process
considered in the previous section could be relevant to the
coarse-grained dynamics of certain driven-diffusive models. To
this end we introduce a simple driven-diffusive model,  and
analyze it using the generalized ZRP model discussed above. This
model is related to a class of models whose coarse-grained
dynamics has been described in terms of the ZRP
\cite{Kafri02A,Kafri03}. In contrast to those models, the model
presented here exhibits also domain diffusion. The model evolves
in discrete time (namely through a parallel update scheme), and is
therefore related to cellular automata models introduced to study
traffic flow \cite{Chowdhury00,Helbing01}.

The model is defined on a lattice of $L$ sites. Each site can
either be vacant ($0$), or occupied by positive ($+$) or negative
($-$) particle. The dynamical moves are carried out in parallel by
two consecutive update steps.
\begin{description}
\item{Step 1: } \vspace{-22pt}\begin{equation*} \eqalign{ x 0 x
&\mathop{\longrightarrow}^{1} 0 x x \nonumber }
\end{equation*}
\item{Step 2: }
 \vspace{-22pt}\begin{equation*} \eqalign{
+- \mathop{\longrightarrow}^{q} -+ \\
+00 \mathop{\longrightarrow}^{a} 0+0 & 00-
\mathop{\longrightarrow}^{a} 0-0 \\
+0000 \mathop{\longrightarrow}^{a} 0+000 \qquad & 0000-
\mathop{\longrightarrow}^{a} 000-0\\
+0000 \mathop{\longrightarrow}^{u} 00+00 \qquad & 0000-
\mathop{\longrightarrow}^{u} 00-00 \;. }
\end{equation*}
\end{description}
Here $x$ is a particle of either type and $u+a \leq 1$. The model
allows for single and double site jumps, and it exhibits features
which do not exist in models with only single-site jumps.  Note
that while all transition probabilities in step 2 are symmetric
under particle-type exchange and left-right reflection, this
symmetry is broken by the transitions of step 1. At high density
the system has states in which all vacancies are isolated, and are
thus deterministic. In this case ergodicity is broken.

\begin{figure}[t]
\centerline{\epsfig{file=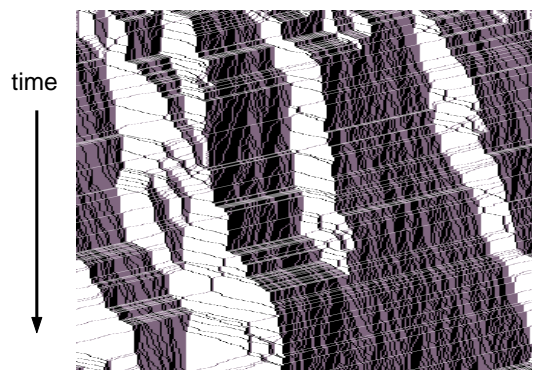,width=7truecm}}
\caption{Space time configurations of the driven model starting
from random initial configuration. A system of size $L=300$ is
occupied by $100$ particles of each type, and the model parameters
are $a=0.45, u=0.5$ and $p=0.1$. Positive particles are colored
black, negative particles are in gray, and vacancies are in white.
\label{fig:config}}
\end{figure}

Consider now the coarse-grained dynamics of this model. Let us
define a domain in this model as a sequence of particles,
interrupted only by isolated vacancies. These vacancies move
deterministically to the left. The evolution of a domain can be
described in terms of two processes: (a) a chipping process, in
which a positive (negative) particle leaves a domain of size $k$
to the right (left) with rate $w_k^+$ ($w_k^-$); and (b) a
diffusion process, in which a vacancy penetrates the domain from
its right (with probability $a$) and advances deterministically to
its left, thus shifting its center of mass one unit to the right.
In figure \ref{fig:config} a space-time configuration of the model
is given. One readily observes the evolution of domains through
chipping and diffusion, and the coalescence of domains upon
contact.

We now consider the current which flows through a domain. The size
$k$ of a domain is defined as the number of particles it contains,
ignoring the vacancies in it. Since vacancies move
deterministically through a domain, one may project the vacancies
out of the domain dynamics, and consider only the dynamics of the
charges. This dynamics is in fact described by the totally
asymmetric exclusion process (TASEP) \cite{TASEP}. Typically, a
domain is asymmetric, composed of an unequal $+$ and $-$ charge
densities. In its maximal current state, obtained for large $a$
and $u$, the current through such a domain takes the form $j_k
\sim 1+b/k$ with $b=3/2$ \cite{Mallick,Sasamoto}. We thus conclude
that $w^\pm_k = w^\pm_\infty\left(1+b/k\right)$, with $w^+_\infty
\neq w^-_\infty$ some constants, and $b=3/2$.

The coarse-grained domain dynamics of this model is thus related
to the generalized ZRP picture of Section \ref{sec:zrpcm}, with
$q\neq\half$ and $b=3/2$. Since our numerical data
(figure~\ref{fig:zrpcm_oned_tau}) suggests that $b^*>3/2$, we
expect the domain size distribution $p_k$ to take the form $p_k
\sim k^{-\tau}$, with $\tau = 2$. Within the domain length
accessible in numerical studies one could not get an accurate
estimate for the exponent $\tau$. However, numerical simulations
of the model indicate that $\tau$ is larger than its
zero-diffusion limit ($3/2$), and is close or equal to $2$. Note
that the correspondence between this model and the zero-range
process can only be made in the region of the model parameters
where the domains are in the maximal-current state, and where the
absorbing states are not reached.

Before concluding let us note that nearest-neighbor attractive
interaction between particle of the same species could result in
larger values of $b$, possibly reaching $b>b^* \simeq 2$. In this
case the dynamics within a domain becomes identical to that of the
KLS model \cite{KLS,Hager01}. Here $w^\pm_k =
w^\pm_\infty\left(1+b(\eta)/k\right)$, where $\eta$ is the
strength of the interaction \cite{Kafri03}. For large values of
$\eta$ one expects the exponent $\tau$ to increase beyond
$\tau=2$. This should be accompanied by a phase separation at high
densities.

\section{Summary and Discussion}
\label{sec:summary}

In this work we generalize a class of zero-range processes with
hopping rates of the form $\omega\left(1+b/k\right)$ to include
diffusion. In particular the exponent $\tau$ associated with the
occupation distribution $p_k \sim k^{-\tau}$, and its dependence
on the parameter $b$, is studied. Both mean-field approximation
and numerical analysis show that for sufficiently small $b$ the
exponent $\tau$ is $b$ independent, and assumes its $b=0$ value.
In particular, in the mean-field approximation $\tau=5/2$ while
for asymmetric nearest-neighbor chipping numerical simulations
indicate that $\tau=2$. Above a critical value of $b$ the exponent
$\tau$ becomes a continuously increasing function of $b$.

This model could be relevant for describing the coarse-grained
dynamics of certain driven diffusive models, and possibly traffic
models. A particular example of a driven diffusive system for
which this is the case is presented.

The fact that $\tau=2$ in the asymmetric chipping-model has been
interpreted as an indication that jamming in traffic models do not
take place via a sharp phase transition \cite{Levine03}. The
analysis presented in this paper indicates that this result is
rather robust, and could hold even for a more general class of
traffic models, for which the chipping-like process exhibits a
weak dependence on the length of the domain ($b<b^*$). Such
dependence may be of relevance for traffic models which allow
overtaking.

\ack

We thank L. Gray for useful discussions. The support of the Israel
Science Foundation (ISF) is gratefully acknowledged.

\end{document}